\documentclass[10pt]{article}
\usepackage{microtype}
\usepackage[margin=1in]{geometry}
\usepackage{lmodern}
\usepackage[T1]{fontenc}
\usepackage{hyperref}
\usepackage{graphicx}
\usepackage{graphicx}
\usepackage{float}
\usepackage{etoolbox}
\usepackage{mathtools}
\usepackage{cleveref}
\usepackage{booktabs}
\usepackage{multirow}

\usepackage{siunitx}
\begin{document}
%\begin{CJK}{UTF8}{gbsn}

\title{The current opportunities and challenges of Web 3.0}
\date{}
\author{Yuqing Fan$^{1,2}$,Tianyi Huang$^{1,2}$,Yiran Meng$^{1,2}$,Shenghui Cheng$^{1,2}$  \\
$^1$ Research Center for the Industries of the Future, Westlake University,Hangzhou,China \\
$^2$ School of Engineering,Westlake University,Hangzhou,China
}

\maketitle 

\begin{abstract}
With recent advancements in AI and 5G technologies,as well as the nascent concepts of blockchain and metaverse,a new revolution of the Internet,known as Web 3.0,is emerging. Given its significant potential impact on the internet landscape and various professional sectors,Web 3.0 has captured considerable attention from both academic and industry circles. This article presents an exploratory analysis of the opportunities and challenges associated with Web 3.0. Firstly, the study evaluates the technical differences between Web 1.0, Web 2.0, and Web 3.0, while also delving into the unique technical architecture of Web 3.0. Secondly, by reviewing current literature, the article highlights the current state of development surrounding Web 3.0 from both economic and technological perspective. Thirdly, the study identifies numerous research and regulatory obstacles that presently confront Web 3.0 initiatives. Finally, the article concludes by providing a forward-looking perspective on the potential future growth and progress of Web 3.0 technology.
\end{abstract}

\section{Introduction}

The Internet began in 1969 in the United States with the AppaNet, initially limited to research departments, schools, and government departments. In 1980, independent business networks began to develop. In 1989, Tim Berners Lee formally proposed the idea of the World Wide Web, and in 1990, he developed the world's first web browser in the European particle physics laboratory in Geneva. In today's world, the Internet has become an indispensable part of production and life, and its development is related to the overall progress of an era. Since its inception, the Internet has gone through three stages: 1) Web 1.0 is an information-connected web, where web pages are static and do not provide interaction; 2) Web 2.0 is the connection between people, where people can communicate on web pages; 3) Web 3.0 is a semantic network. In addition to interaction, Web 3.0 also covers a large amount of data. Due to its decentralized characteristics, it can freely access data resource \cite{hiremath2016}.

Web 3.0 was proposed by John Markoff in the New York Times in 2006. Research has pointed out that Web 3.0 is a symbiotic entity of web technology and knowledge representation, a subfield of artificial intelligence \cite{hendler2009web}. Some studies also suggest that Web 3.0 is the next-generation network that integrates technology, law, and payment systems\cite{ref10}. With the continuous development of Web 3.0, it has integrated many emerging concepts, such as the metaverse, digital economy, and 5G. This integration is constantly changing people's lives. For example, people can immerse themselves in the places they want to go home through virtual reality (VR) technology and the internet and interact with their peers' virtual digital human images. These changes have made people's expectations for the internet increasingly strong. Table \ref{对比} compares the differences between Web 1.0, Web 2.0, and Web 3.0. From Web 1.0 to Web 3.0, interaction has improved, and the amount of information available has also significantly increased. This continuously improves the user experience\cite{cheng2022roadmap}.

\begin{table}
\centering
\renewcommand{\tablename}{Table}
\caption{Comparison among Web 1.0,Web 2.0 and Web 3.0}
\Large
\scalebox{0.7}{
\begin{tabular}{|c|c|c|c|}
\hline
 & \textbf{Web 1.0}  & \textbf{Web 2.0}   & \textbf{Web 3.0}   \\ \hline
\textbf{Time}   & 1996 & 2006 & 2016 \\ \hline
\textbf{Attribute} & Hypertext & Social media & Semantic web \\ \hline
\textbf{Medium} & Static texts & Interactive contents & Virtual contents\\ \hline
\textbf{Infrastructure} & PC & Cloud and mobile device & Blockchain  \\ \hline
\textbf{Interaction}   & \begin{tabular}[c]{@{}c@{}} Company publishes, \\ users can only read  \end{tabular}  & \begin{tabular}[c]{@{}c@{}} Company establish platforms \\ for users to interact \end{tabular} & \begin{tabular}[c]{@{}c@{}} Anyone can \\ establish platforms \end{tabular}   \\ \hline
\textbf{Searching}&\begin{tabular}[c]{@{}c@{}} Widely search, \\ results are very vague \end{tabular} &\begin{tabular}[c]{@{}c@{}} Search with keywords \\ bring with accurate results \end{tabular}&\begin{tabular}[c]{@{}c@{}} More accurate results \\ using big data \end{tabular}\\ \hline

\end{tabular}}
\renewcommand{\tablename}{Table}
\label{对比}
\end{table}

\begin{figure}[H]
    \centering
    \includegraphics[scale=0.75]{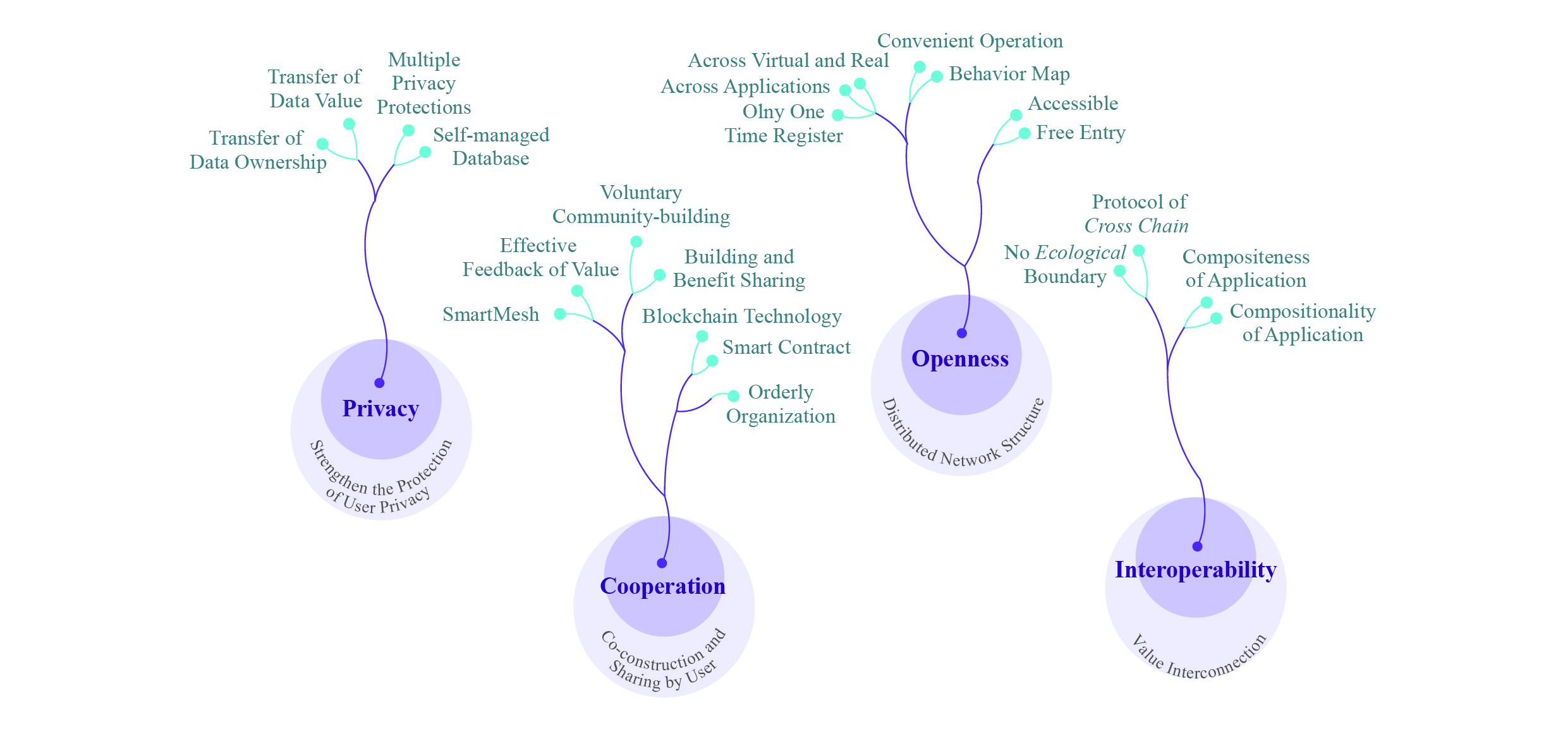} 
    \renewcommand{\figurename}{Figure}
    \caption{4 characteristics of Web 3.0}
    \label{Web 3.0的四个特点}
    
\end{figure}

The emergence of Web 3.0 represents a significant evolution in the realm of Internet technologies, marked by four key characteristics distinguishing it from its predecessors. Firstly, Web 3.0 is characterized by openness, which enables users to access various platforms utilizing only one account. Secondly, data privacy is ensured via the decentralized structure of blockchain technology that protects user data ownership and eliminates reliance on third-party management platforms. The third key feature of Web 3.0 relates to cooperation, which is facilitated through token incentives designed to reward content creators for their contributions and foster a more equitable platform. Lastly, interoperability refers to the degree to which third parties no longer constrain user behaviors, thereby offering greater latitude in managing personal activities and engagements across varying digital environments. See figure \ref{Web 3.0的四个特点} for details.

Web 3.0 has garnered significant attention, with numerous high-tech companies rapidly positioning themselves within this emerging landscape. For instance, Twitter has begun integrating various Web 3.0 applications into its products, while Google established a dedicated Web 3.0 team in May 2022. Moreover, many notable celebrities, including Stephen Chow, Jay Chou, and Junjie Lin, have prominently featured NFTs among their holdings. Importantly, at the national level, Web 3.0 represents a novel field of competitive engagement between countries. This competition extends beyond technological prowess, entailing a contest for influence in the unfolding Internet ecosystem. The implications of these developments for the future of digital society are profound and warrant further examination and exploration by scholars and practitioners alike.

In conclusion, Web 3.0 constitutes a profound transformation that will significantly impact various aspects of our social, technological and economic lives. This comprehensive article aims to provide a detailed introduction to the concept of Web 3.0. The contributions of this paper are multi-fold, including: 1. a thorough exposition of the connotations and applications of Web 3.0, and a comprehensive analysis of its current development status from the perspectives of policy, market, and scientific research; 2. an exploration of the changes and opportunities that Web 3.0 presents, identifying its developmental requirements based on these changes and opportunities; 3. an overview of the future prospects for the evolution of Web 3.0, addressing the potential risks emanating from this new technological paradigm.

\section{Connotation and application}

This section first introduces the technology stack of Web 3.0, then provides six common functions of technology stack and provides use cases.

\subsection{Web 3.0 technology stack}

As of January 2023, many papers have listed the technology stacks of Web 3.0. Guan \cite{guan2022web3} cited the five-layer technology stack proposed by the Web3 Foundation in 2017. The zero layer includes node communication and low-level programming. The first layer is responsible for providing interaction and transferring data. The second layer is used to extend the functions of the first layer. The third layer provides a readable language for development. Ordinary users interact in the fourth layer. Jacksi \cite{jacksi2019development} proposed a four-layer technology stack for Web 3.0. The first layer is the unified resource identifier, the second layer is the hypertext transfer protocol(HTTP), the third layer is the file transfer protocol, and the last layer is the IP security protocol. Swati \cite{swati2022blockchain} gave a Web 3.0 technology stack composed of the application layer, blockchain framework layer, and blockchain network layer. In addition, Petcu \cite{petcu2023secure} also proposed different Web 3.0 technology stack architectures. Rasekh \cite{rasekh2012dynamic}, Sarchandraa \cite{sarathchandra2021decentralized} and Zheng \cite{zheng2020ethereum} proposed the technology stacks of Semantic Web, Ethereum, and DApp respectively.

The Web 3.0 technology stack proposed by this paper is shown in \ref{jsz}, which is divided into six layers:

%本文提出的Web 3.0技术栈如\ref{jsz}所示，共分为六层：

\begin{figure}[H]
    \centering
    \includegraphics[scale=0.5]{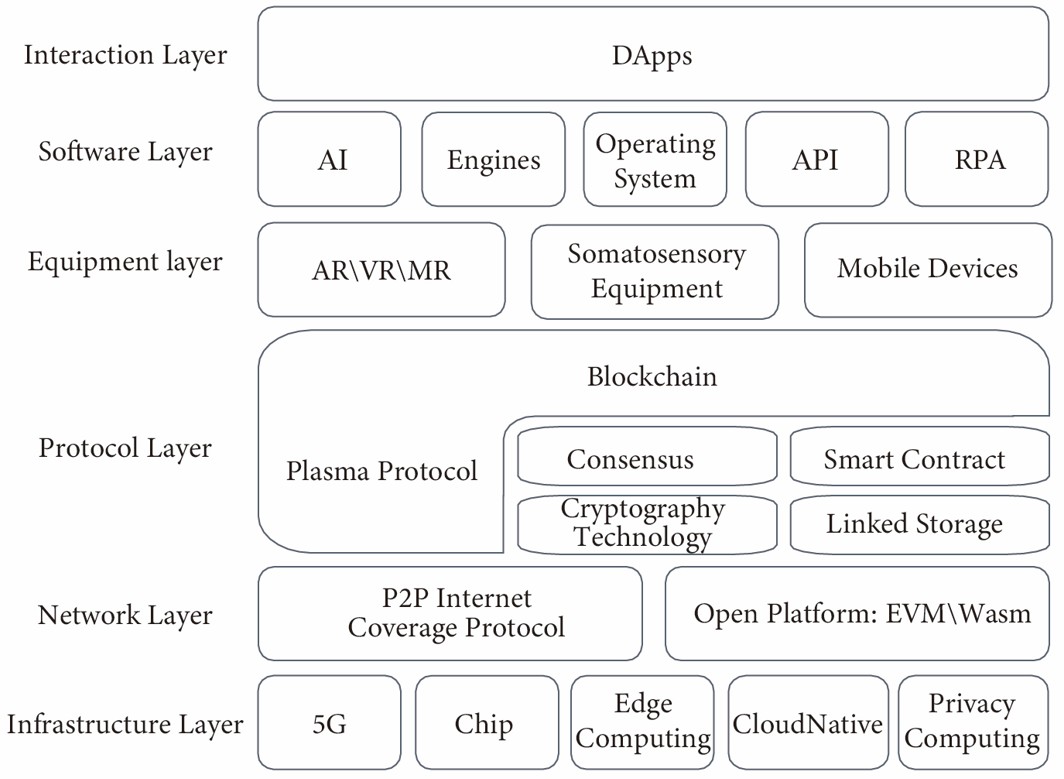} 
    \renewcommand{\figurename}{Figure}
    \caption{Web 3.0 technology stack}
    \label{jsz}
\end{figure}

At the bottom of the stack is the infrastructure layer. It provides a faster network environment for Web 3.0. 5G technology adds the file form that can be transferred \cite{23}; Chips expand the overall computing power of the device; The distributed architecture of edge computing alleviates traffic congestion \cite{24}; Cloud native is used to help software development; Privacy computing can effectively prevent information disclosure\cite{26}.

Infrastructure development can drive the development of the network layer and the protocol layer. For example, P2P protocol can enable efficient file transmission and greatly relieve the pressure on the server. Plasma protocol improves the throughput of blockchain in the form of "The chain in the chain"\cite{14}. In addition, there are consensus mechanisms that allow nodes to vote, cryptography technologies that enhance security on the chain, smart contracts that allow no third-party transactions, and chained storage.

Through various expansion devices, the equipment layer of Web 3.0 can provide users with multi-dimensional experiences, such as touch and hearing. For example, AR/VR/MR and other devices that bring 3D experience provide tactile changes in body sense devices and mobile devices that make applications more convenient.

Above the equipment layer is the software layer. Robotic Process Automation (RPA) can realize the automation of massive repetitive processes and reduce labor costs. AI has added a learning function to RPA for intelligent operation\cite{25}. As the core of the development program, the engine can help developers quickly lay the program. The operating system provides management capabilities for terminal devices and software. API can help software interact well with other software.

This is how DApps are built on the technology stack. Compared with traditional software, DApps have made great progress in a distributed architecture, body-sensing devices, software collaboration, and other aspects. As a result, they can bring users a cross-era experience.

\subsection{Web 3.0 applications}

In recent years, the applications of Web 3.0 has been in a blowout state. According to a survey, 57\%  of respondents experienced at least one Web 3.0 application in 2022. By January 2023, the number of DAPPs in the global market has reached 6370, and the number of smart contracts has reached 16140~\cite{21}. Table \ref{功能} shows the  classification of typical Web 3.0 applications on the market at present:

\begin{table}
\centering
\centering
\renewcommand{\tablename}{Table}
\caption{Classification of Web 3.0 applications}
\setlength{\tabcolsep}{1mm}{
\begin{tabular}{|c|c|c|}
\hline
Classification & Description & Examples \\ \hline
Finance & \begin{tabular}[c]{@{}c@{}} Issue,trade and manage financial products \\ and other traditional services \end{tabular}  & Uniswap,MetaMask \\ \hline
Infrastructure & \begin{tabular}[c]{@{}c@{}} Provide infrastructure services \\ such as blockchain or on-chain storage \end{tabular} & Ethereum,Antgroup \\ \hline
Social & \begin{tabular}[c]{@{}c@{}} Digital community composed of enthusiasts \\ also contain media functions \end{tabular} & Sapien,Steemit,Meta \\ \hline
Multimedia & \begin{tabular}[c]{@{}c@{}} Provide a variety of creative methods, \\ and ensure creators can obtain \\ corresponding benefits \end{tabular} & BanklessDAO, Forefront  \\ \hline
Game & A new form of game & Crypto Dynasty,Splinterlands  \\ \hline
Service & \begin{tabular}[c]{@{}c@{}} Other services provide to users \\ including telecommuting and digital medicine \end{tabular} & Horizon Workrooms,Ethlance  \\ \hline
\end{tabular}
}
\label{功能}
\end{table}

The financial industry is the field where many Web 3.0 applications are based. Web 3.0 can be combined with existing financial instruments to create a new business model\cite{16}. Ozili \cite{17} et al. divided the existing Web 3.0 financial applications into nine types: stable currency levy, loan, NFT issuance, non-intermediary transaction, secondary market transaction, liquidity, e-wallet, and asset management. The digital banking software released by banks automates traditional banking services through intelligent devices, bringing great convenience to customers.

Some enterprises choose to rent or sell infrastructure services and platforms related to Web 3.0. Taking blockchain as an example, many enterprises in China, such as Ant, Tencent, and Baidu, have provided alliance chain solutions; Onchain.storage chooses to provide on-chain storage services to help other DApp developers provide data storage space.

Social media has been very hot in the era of Web 2.0, and the arrival of Web 3.0 will further upgrade communication between people. Meta has brought the metauniverse technology into social interaction and brought participants an immersive experience through virtual imaging. Applying VR and somatosensory devices will also bring further innovation to the social model.

The multimedia of Web 3.0 includes two specific functions. The first is providing creators with as many creative methods as possible, and the second is establishing a platform to ensure that creators can obtain corresponding benefits. Some digital design platforms even allow customers to customize their large-scale functions\cite{30}. The application of this function includes RaidGuild, which provides consulting, design, development, and marketing services—Flickr, which creates and shares diversified photos.

The virtual nature of the game makes it the first to embrace the ecosystem of Web 3.0. As a combination of game and decentralized finance, chain game allows assets in the game to circulate in the real world as commodities. Decentraland is a 3D digital game platform. Players can exchange collectibles, purchase and sell digital assets in the game, and also interact through wearable devices \cite{28}.

In addition to the above five functions, there are some scattered functions. This article integrates it into other services that can provide users with an experience. It includes remote office service software such as Horizon Workrooms and Ethlance, Web 3.0 browsers with built-in encrypted wallets such as Brave and Osiris, and digital medical services for managing hospital files and assisting with medical examinations.

Of course, most Web 3.0 applications have multiple functions simultaneously. For example, financial services and social tools are a common combination. Take MakerDAO as an example, and its participants can modify the Maker agreement by holding MKR tokens to vote. The MakerDAO community also has a stable currency Dai for storage, exchange, payment, and bookkeeping. The community also provides blogs and forums for participants to conduct social activities \cite{18}.

%------------------------------------------%

\section{Development status}

The development of theory is a prerequisite for the development of Web 3.0 applications. Web 3.0 as a research area has a history that dates back to 2006. Although still in its nascent stage, significant strides have been made towards its advancement. For example, blockchain technology has found widespread application in the financial sector, whereas intelligent contracts and distributed applications are continuously being developed and promoted. Other fundamental components of Web 3.0 such as semantic Web, distributed file systems, metadata application, and the Internet of Things (IoT) are in the process of implementation or are developing towards implementation, arousing extensive industry interest. Enterprises recognize that Web 3.0 technology will significantly transform future business landscapes. Based on diverse perspectives from home and abroad, and buttressed by empirical data, this paper examines the status of Web 3.0 from three dimensions of market, policy, and scientific research, analyzing its developmental trajectory and charting possible future directions.

\subsection{Market status}

\begin{figure}[H] 
    \centering
    \includegraphics[scale=0.5]{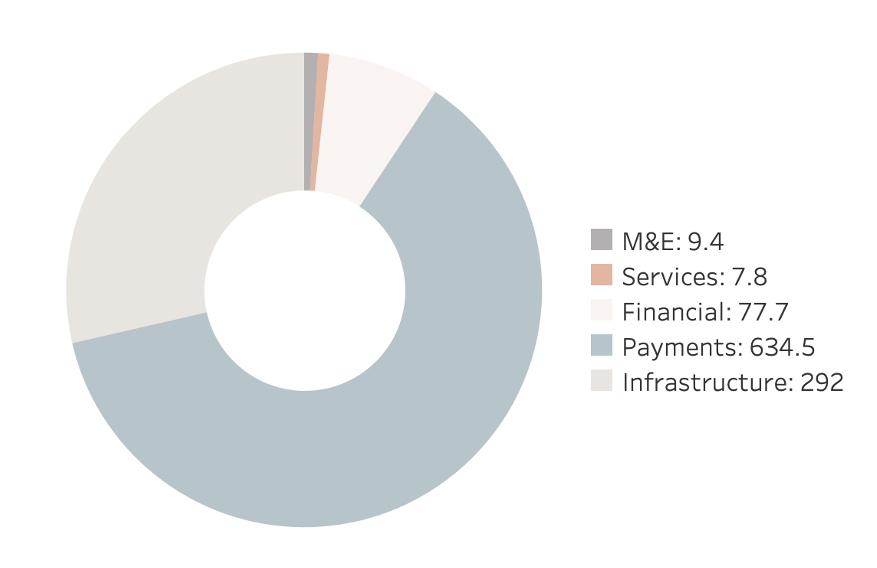} 
    \renewcommand{\figurename}{Figure}
    \caption{Web 3.0 market scale. Unit: billion dollars}
    \label{市场规模}
\end{figure}

In 2014, the founder of Ethereum described the ideal operation mode of Web 3.0. By 2020, with the rapid spread of the concept of the metauniverse, Web 3.0 will arouse public discussion for another time. In recent years, the market size of Web 3.0 has been increasing. This paper uses Messari \footnote{Messari is the most prominent Web 3.0 research institute at present. The institute provides real-time data and research reports} of Web 3.0-related industries to preliminary analyze the current market size. Figure \ref{市场规模} shows the market size of various fields under the current Web 3.0 ecosystem. By January 29, 2023, the market size of Web 3.0 had reached US \$102.14 billion, of which the largest market size was in the field of payments, reaching US \$63.45 billion. On the other hand, the market size of Media and Entertainment is the smallest, with only US \$940 million.

\begin{figure}[H] 
    \centering
    \includegraphics[scale=0.9]{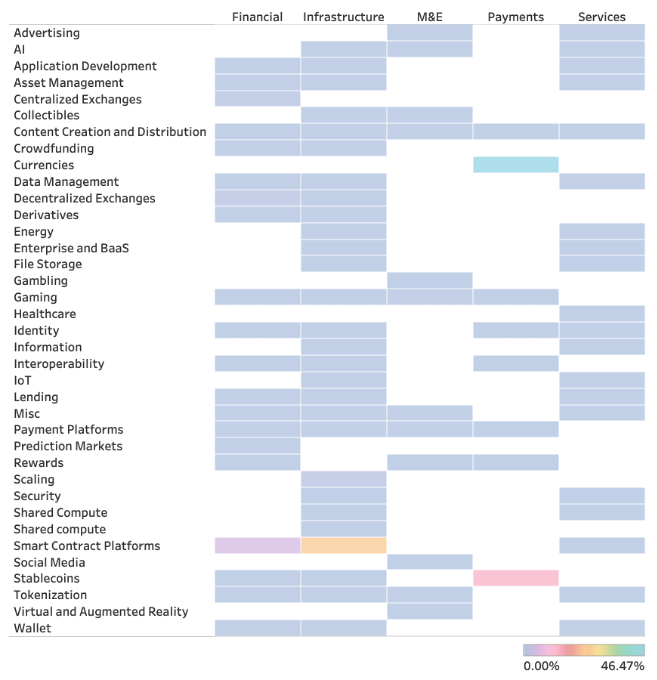} 
    \renewcommand{\figurename}{Figure}
    \caption{Web 3.0 market size contribution of each category: the horizontal axis is domain classification, and the vertical axis is business classification. White indicates that there is no corresponding business under the field}
    \label{贡献度}
\end{figure}

Figure \ref{贡献度} uses the thermal diagram to analyze the contribution of different businesses in various fields of the Web 3.0 ecosystem. It can be seen from the figure that the Currency business in the Payment field is the largest, about 46.47\%. This number shows that the number of companies in the payment field is significant. Then is the Smart Contract Platform business in the infrastructure field, the Stablecoins business in the payment field, and the Smart Contract Platform business in the financial field. Through the analysis of market scale contribution, we can see that the major development of the Web 3.0 market comes from electronic payment. The digital financial industry based on blockchain technology has begun to take shape. However, other fields are still in the early stage of development with significant room for growth.

Unlike the international market, China's development has adopted the path of alliance chain as the primary way and adopted the separation strategy of chain and currency to reduce the Internet financial credit risk under blockchain technology as much as possible. According to the statistics of China Information and Communication Academy, the revenue scale of China's alliance chain business in the first half of 2022 will reach about 2 billion yuan. The development of the alliance chain has also promoted the growth of domestic blockchain as a service (BaaS) business. In 2021, China's BaaS business market will reach 188 million US dollars, with a growth rate of 92.6\%. Domestic Internet companies, blockchain technology service providers, universities and research institutes, and other joint research and development alliance chains are gradually being established \cite{22}. At present, leading domestic technology enterprises, such as Tencent, Alibaba, and Huawei, have invested a large amount of research and development funds in the technical fields related to Web 3.0, actively promoting enterprises to use the Web 3.0 ecosystem to realize digital transformation, which has laid a technical foundation for the development of Web 3.0 in China.

\subsection{Policy status}

Currently, the most important national laws and regulations on Web 3.0 are mainly focused on virtual currency. The government has issued several laws and regulations on virtual currency and its derivatives related to Web 3.0. Table \ref{监管} shows some of the current regulatory policies of the country. It can be seen that the focus of China's current regulation is how to regulate the malignant transfer of legal currency to virtual currency, how to regulate the malicious injection or withdrawal of foreign capital into the domestic market through virtual currency, and how to prevent speculators from using Web 3.0, blockchain, NFT and other emerging concepts to make malicious speculation and create economic foam\cite{ref10}.

\begin{table}
\centering
\renewcommand{\tablename}{Table}
\caption{Some Chinese national regulatory policies on Web 3.0 related fields}
\begin{tabular}{|c|c|}
\hline
Field  & Regulatory policies \\ \hline
\multirow{3}{*}{Virtual currency} & Notice on protection against Bitcoin risks \\ \cline{2-2} 
& \begin{tabular}[c]{@{}c@{}} Announcement on preventing the risk \\ of token issuance financing \end{tabular} \\ \cline{2-2} 
& \begin{tabular}[c]{@{}c@{}}Notice on Further Preventing and Dealing with \\   the Risks of Fictitious Currency Trading \end{tabular}\\ \hline
Blockchain  & Blockchain information service management regulations  \\ \hline
Meta-universe & \begin{tabular}[c]{@{}c@{}} Risk tips on preventing illegal fund-raising \\ in the name of "meta-universe" \end{tabular} \\ \hline
NFT  & Initiatives on preventing NFT-related financial risks  \\ \hline
\end{tabular}
\label{监管}
\end{table}

In addition to regulating virtual currencies, state authorities are required to oversee a wide range of aspects concerning the development and utilization of Web 3.0 technology. While strict regulations may mitigate potential financial risks, they also have the potential to impede domestic technological innovation. Consequently, it is necessary to consider alternative means of preventing and reversing any trend towards a shift from reality to virtual brought about by Web 3.0. Equally important is the prevention of malicious forms of competition amongst internet enterprises stemming from changes to business models brought on by Web 3.0 technology. Hence, governments need to adopt more forward-looking and innovative regulatory strategies to promote sustainable long-term growth in this area.

\subsection{Research status}

This study offers empirical evidence to facilitate a comprehensive analysis of the current state of international research on Web 3.0 technology by surveying relevant documents indexed in the Web of Science database. Furthermore, the Chinese knowledge infrastructure (CNKI) is employed to examine the corresponding status of domestic research. By conducting a comparative analysis of the relevant scholarship on Web 3.0 in China and abroad, this paper enables an investigation of any notable differences between these two research contexts.

\begin{figure}[H] 
    \centering
    \includegraphics[scale=0.4]{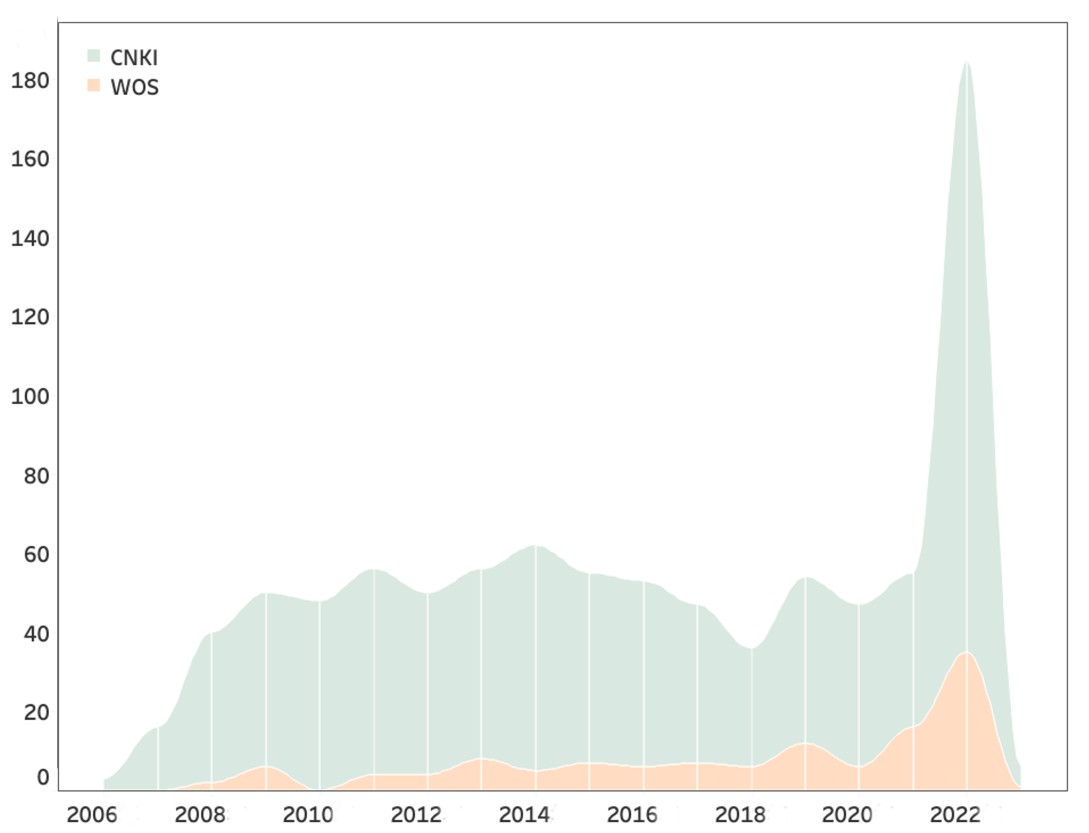} 
    \renewcommand{\figurename}{Figure}
    \caption{Web 3.0 related literature publication, 2003-2023}
    \label{5a}
\end{figure}

\begin{figure}[H] 
    \centering
    \includegraphics[scale=0.5]{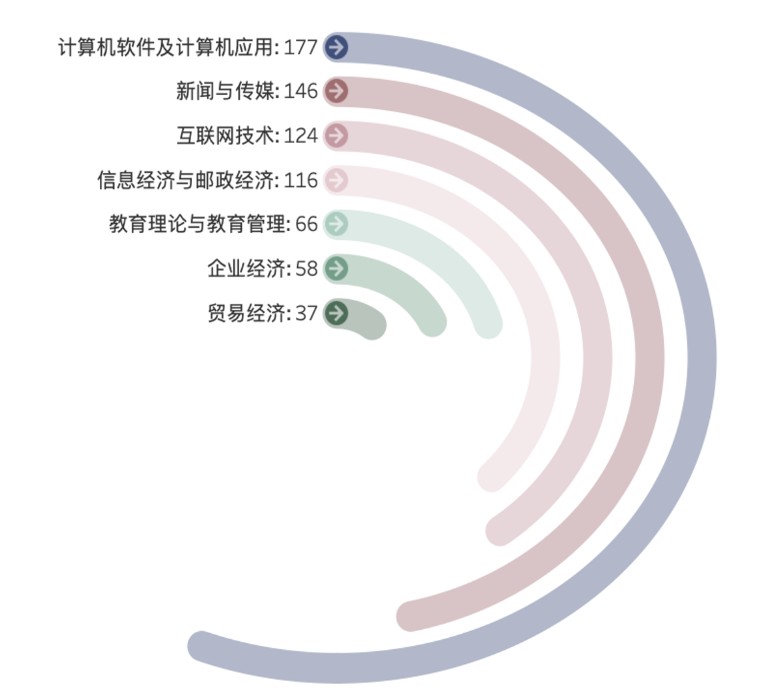} 
    \renewcommand{\figurename}{Figure}
    \caption{Distribution of subjects of Web 3.0 related literature in CNKI database. From most to least: Computer software and applications;News and Media;Internet technology;Information and Post economy;Education theory and management; Entrepreneur economy;Trade economy}
    \label{5b}
\end{figure}

\begin{figure}[H] 
    \centering
    \includegraphics[scale=0.5]{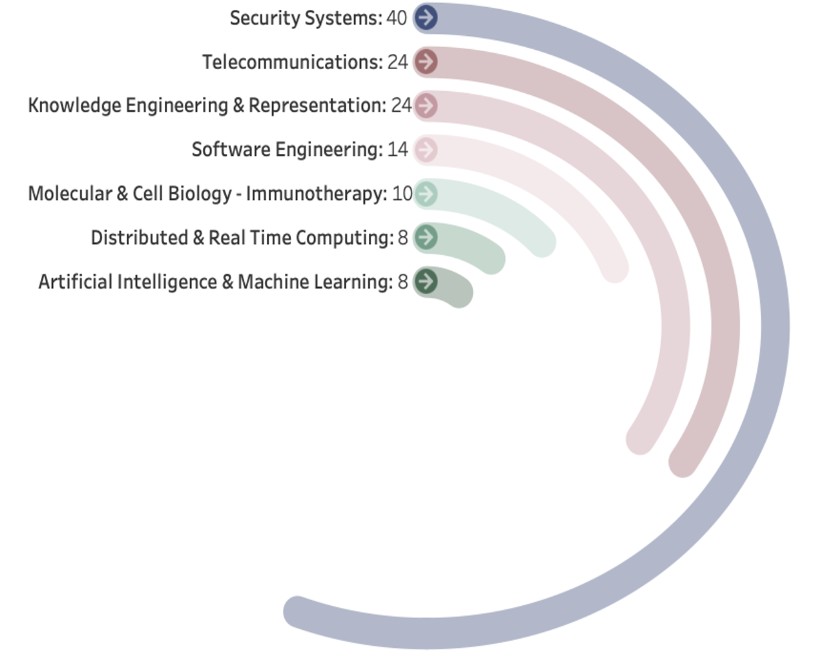} 
    \renewcommand{\figurename}{Figure}
    \caption{Distribution of subjects of Web 3.0 related literature in WOS database}
    \label{5c}
\end{figure}

As of January 29, 2023, 125 documents with high relevance to Web 3.0 have been retrieved in the WOS database, while 794 documents related to Web 3.0 have been found in the CKNI database. Figure \ref{5a} shows the changing trend of the number of documents related to Web 3.0 published at home and abroad. In 2006, research related to Web 3.0 began to appear in China. After reaching the first peak in 2014, the number of documents gradually decreased, and until 2020, the number of related documents increased sharply. Although the overall number of foreign documents related to Web 3.0 is less than that in China, its change trend is similar to that in China. Figures \ref{5b} and \ref{5c} show the disciplinary distribution of relevant literature, respectively. Domestic research mainly focuses on computer software and computer applications, news and media, and Internet technology, while foreign research focuses on security systems and telecommunications.

In addition, this paper extracts keywords from Web 3.0-related literature to analyze the research hot spots in this field. The hot topics in the past literature represent the research focus of this field in a certain period, and the frequency of keyword occurrence is the specific manifestation of the research focus. This paper carries out keyword clustering analysis and emergent keyword analysis, providing references for researchers to explore the iterative development and frontier trends of Web 3.0 hot topics.

\subsubsection{Keyword trending among Chinese literature}

\begin{figure}[H] 
    \centering
    \includegraphics[scale=0.5]{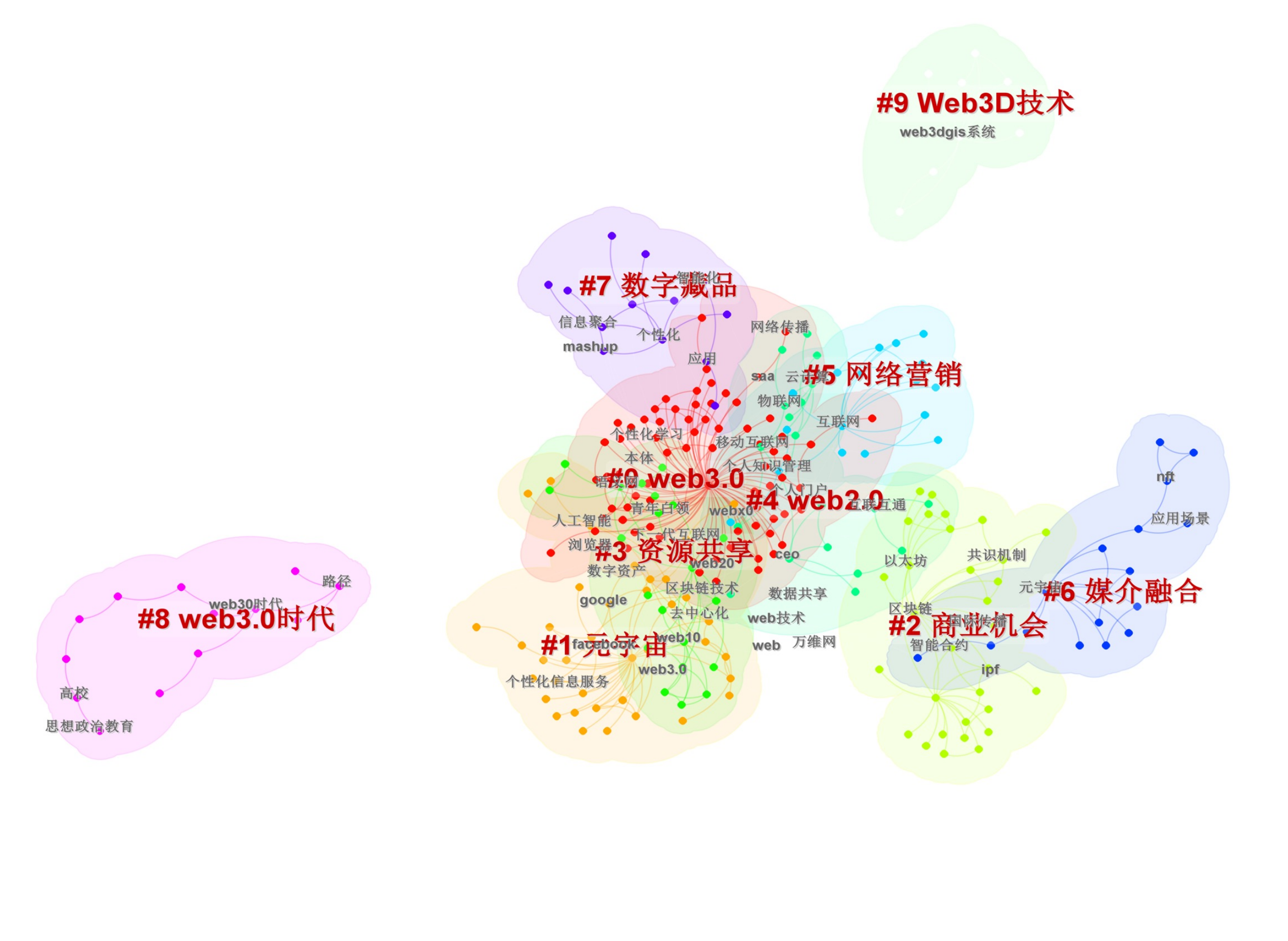} 
    \renewcommand{\figurename}{Figure}
    \caption{CNKI keyword cluster map. Category from 0 to 9:Web 3.0;Metaverse;Business opportunities; Resource sharing;Web 2.0;Internet marketing;Media fusion;Digital collections;The era of Web 3.0;Web 3D technology;}
    \label{ccluster}
\end{figure}

Figure \ref{ccluster} shows the results of keyword clustering analysis for published web 3.0 related papers in China. Based on the clustering analysis results, it can be seen that since 2006, the Web 3.0 literature published in China can be roughly divided into three categories.

The first category is theoretical research, which mainly studies the fundamental theories behind the development of Web 3.0. For example, some researchers discussed the current development trend of Web 3.0 from four aspects: network and computing technology, security and trustworthiness technology, virtual real fusion technology, and intelligent interaction technology; Li Jie \cite{lj2023} believes that the core of Web 3.0 is the Semantic Web, and discusses the relationship between Web 3.0, social 5.0 and the metauniverse; Wu Tong et al. \cite{wt2023} proposed that the core concept of Web 3.0 is that users have access to data, which enables it to incorporate data into the framework of production factors; Chen Yongwei \cite{16} discussed four possible challenges that Web 3.0 may bring from the perspective of policy designation, and proposed some corresponding measures. In addition, the relevant literature in this category mainly focuses on the market opportunities and value that Web 3.0 may bring. For example, the "Business Opportunities" cluster includes "Business Opportunities", the "Media Integration" cluster includes theme words such as "Model Innovation", and the "Web 3.0 Era" cluster also includes theme keywords related to macro development such as "Collaborative Development" and "Regional Development." However, Web 3.0 is not accurately defined in the theoretical literature. Therefore, from a research perspective, there is currently no unified definition for it.

The second category is technical research, which mainly focuses on the technologies required for the development of Web 3.0, including high-frequency keywords in "resource sharing" clustering that include "mashup technology", and "metaverse" clustering that includes topics such as "blockchain", "smart contracts", "IPFS", and "Truffle framework". In addition, the "Web3D technology" clustering includes keywords such as "3D annotation" and "VRML". These topics are hot topics in the research of Web 3.0 technology. For example, Yun Jian et al. \cite{yj2022} designed a digital archive storage platform based on alliance chain and IPFS technology, which can expand storage capacity while also taking into account data security; Sun Yushan et al. \cite{sys2023} proposed a data structure suitable for efficient authentication of streaming data on the chain, which solves the problem of server untrustworthiness when data is outsourced for storage; Lu Jiawei et al. \cite{ljw2022} discussed a mashup service clustering method based on non-negative matrix factorization combining tags and word embedding (TWE-NMF) topic model, which is more efficient than traditional methods in clustering accuracy. Tianyi Huang et al. \cite{huang2023high} introduced a new data analysis and visulization method based on high-dimensional data, which can be implemented under multiple categories.However, the literature in this category currently only focuses on the technology itself, with almost no specific application of technology in Web 3.0.

The third category is application research, which mainly explores specific applications in the Web 3.0 ecosystem. For example, high-frequency topics such as "online teaching platforms", "electronic libraries", and "NFTs" are included in the "resource sharing" cluster. Tian Linan et al. \cite{tln2022} analyzed the application of spatiotemporal encoding technology in Web 3.0 digital finance using Qingdao City as a model and concluded that spatiotemporal encoding technology could play an essential role in platform traffic distribution, financial risk control, financial supervision, and other scenarios; Zhai Xuesong et al. \cite{zxx2022} analyzed the possible applications of Web 3.0 in education and believed that Web 3.0 could seize the opportunities of major educational strategies such as high-quality personalized teaching, educational equity, and the simultaneous development of "five educations"; Zhao Yichen et al. \cite{zyc2021} proposed using blockchain technology to solve the problem of strenuous tax activities in the digital economy and reduce the phenomenon of tax evasion. From the current literature, research on Web 3.0 application scenarios mainly focuses on online teaching, metaverse, and online marketing. Moreover, more involvement in applied research in other fields must be needed.

Other researchers have integrated the three types of research mentioned above to study the future development of Web 3.0. For example, the book "Web 3.0: A Disruptive and Significant Third Generation Internet" provides a detailed introduction to the basic theoretical knowledge and applications of Web 3.0, and provides suggestions on how to formulate network behavior guidelines in the Web 3.0 era based on literature review\cite{ref10}.

\begin{figure}[H] 
    \centering
    \includegraphics[scale=0.55]{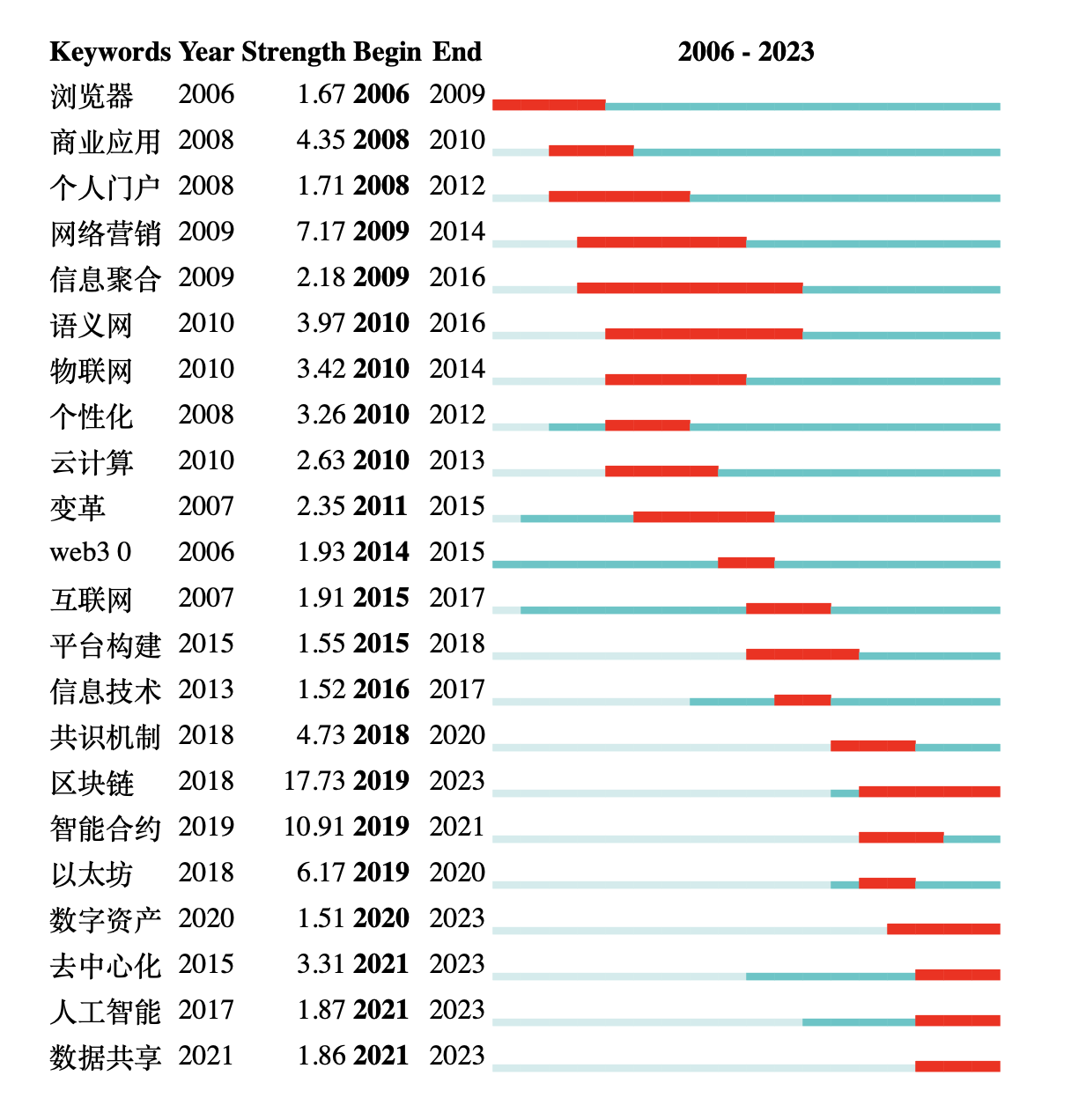} 
    \renewcommand{\figurename}{Figure}
    \caption{CNKI emergent keywords Top20: light green is the basic time axis, dark green is the time interval from the emergence of keywords to the present, and red is the time interval of the duration of keyword emergent. The keywords in the figure are sorted by the starting time of emergent.Keywords from top to below:Browser;Business app;Personal portal;Web marketing;Data aggregation;Semantic Web;IoT;Personalization;Cloud Computing;Transformation;Web 3.0;Internet;Platform construction;IT;Consensus mechanism;Blockchain;Smart contract ;Ethereum;Digital assets;Decentralization;AI;Data sharing}
    \label{cburst}
\end{figure}

Figure \ref{cburst} shows the top 20 keywords. The first emergent keyword was "browser", which emerged from 2006 to 2009, followed by "business application", "personal portal", and "online marketing". This shows that in the early research of Web 3.0 in China, the main direction is to explore new Internet marketing methods. The key word with the most significant strength is "blockchain", and its strength reaches 17.73. In addition, domestic research has gradually shifted from the initial "commercial application" and "online marketing" to the technical topics of "platform construction", "consensus mechanism", and "blockchain". It is worth noting that "digital assets", "decentralization", "artificial intelligence," and "data sharing" have been new research hot spots in the past three years.

\subsubsection{Keyword trending in foreign countries}

\begin{figure}[H] 
    \centering
    \includegraphics[scale=0.55]{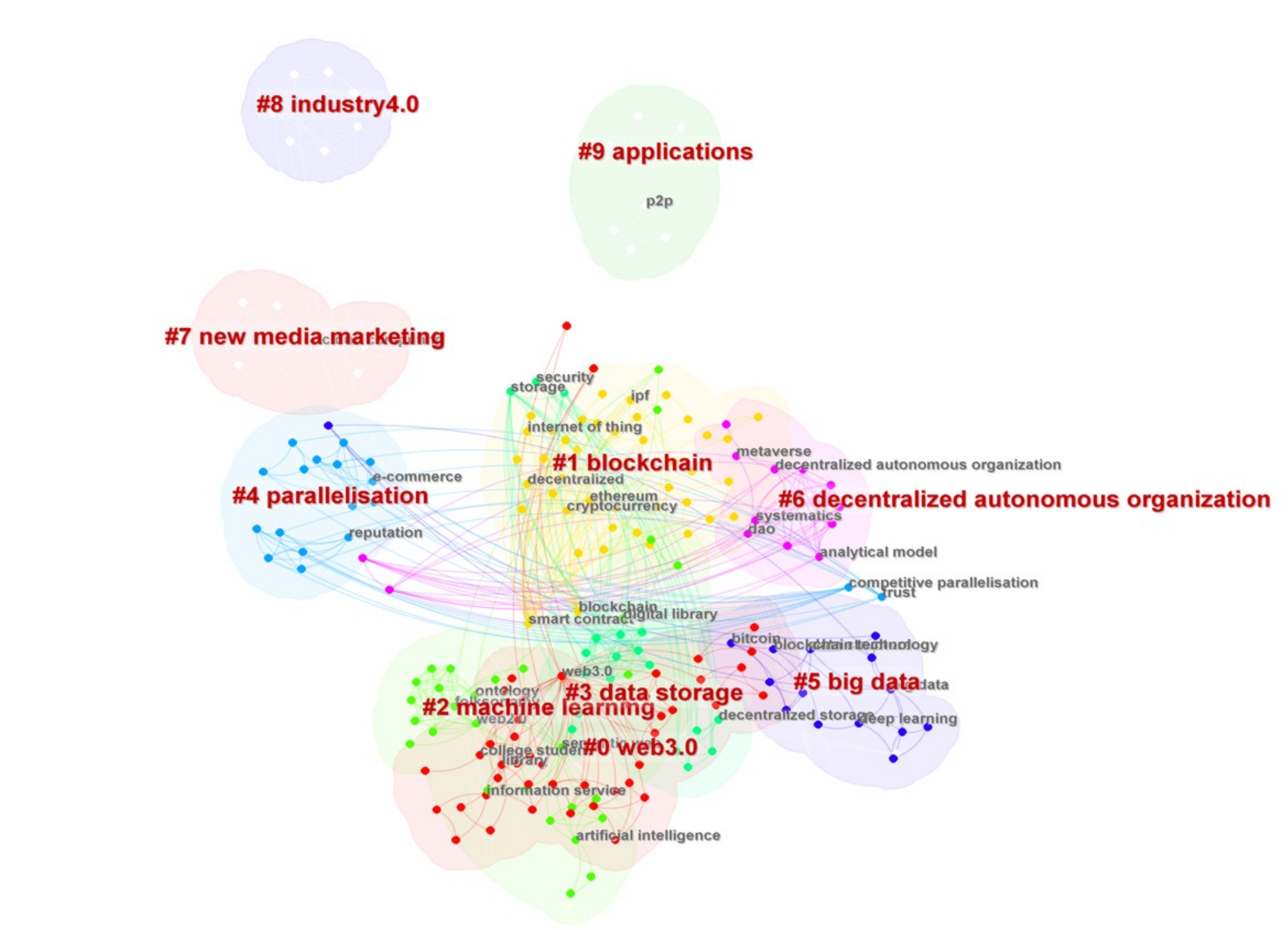} 
    \renewcommand{\figurename}{Figure}
    \caption{WOS keyword cluster map}
    \label{wcluster}
\end{figure}

Figure \ref{wcluster} shows the keyword clustering analysis graph of web 3.0-related literature published abroad. Unlike domestic research, the hot topics related to Web 3.0 research abroad can be divided into four categories: theoretical research, technical research, industrial research, and application research.

The first category is theoretical research. In addition to outlining the overall development, there are also studies abroad that jointly analyze Web 3.0 and social sciences. For example, Rudman and Bruwer \cite{Rudman2016DefiningW3} summarized the overall development and current challenges of Web 3.0; Kreps and Kimppa \cite{Krep2015} discussed the specific definition of Web 3.0 from the perspective of post-structuralism, and demonstrated the social development brought by Web 3.0 through five papers; Ferraro et al. \cite{FERRARO2023} starting from the development of psychology and technology, discussed how to establish trust mechanisms in Web 3.0 and concluded that there are loopholes in the current trust mechanisms.

The second category is technical research, which is similar to domestic research and explores supporting Web 3.0 technology development. For example, in the "Machine Learning" cluster, topics such as "ICA Algorithm" appear, while in the "Data Storage" cluster, keywords such as "Storage Systems" and "IPFS" are included. In addition, Clarke et al. \cite{Clarke2018} proposed that the decentralized communication layer in the mainframe can be used as a supplementary protocol to strengthen the network security of the Internet; Vega et al. \cite{Miguel2022} designed a new cryptographic primitive, Nil Message Compute (NMC). Because nodes under the NMC structure do not run immutable ledgers that store transaction data, nor are they interconnected, it will be more secure than blockchain structures. However, unlike domestic research, there is only a little research on Web3D technology in foreign literature related to Web 3.0.

The third category is industrial research, which mainly focuses on how Web 3.0 integrates with existing industries to build a new industrial ecosystem. Among them, the "Industry 4.0" cluster contains keywords such as "IoT" and "industrial internet of things", and the "Parallelism" cluster contains keywords such as "Supply Chain". Anwar \cite{Anwar2022ASO} explored the connection between Web 3.0 and the Internet of Things (IoT), believing that the collaboration between Web 3.0 and the Internet of Things will benefit society in multiple directions such as smart cities and soil observation; Castro et al. \cite{CASTRO2013396} delved into the connection between Web 3.0 and the manufacturing industry and introduced a virtual enterprise framework specifically designed for small and medium-sized enterprises using the Web 3.0 platform. This framework will help small and medium-sized enterprises adapt faster to changes in industrial structure and achieve sustainable development goals. Compared to domestic research, foreign research has already discussed building a Web 3.0 industry ecosystem.

The fourth category is applied research. Foreign Web 3.0 application research covers many fields, including office, healthcare, education, commerce, economics, and other scenarios. For example, Gupta et al. \cite{Gupta20223DNA} developed a 3D conferencing application in the metaverse and VR devices. This program ensures meeting efficiency and reduces the impact of delays; Ayoade et al. \cite{Ayoade2018} proposed a decentralized data management system suitable for IoT devices. As it does not require central system management, it can better protect data; Thrul et al. \cite{Thrul2022} believe that DAO can provide remote mental health services and solve the problem of scarce mental health resources.

\begin{figure}[H] 
    \centering
    \includegraphics[scale=0.55]{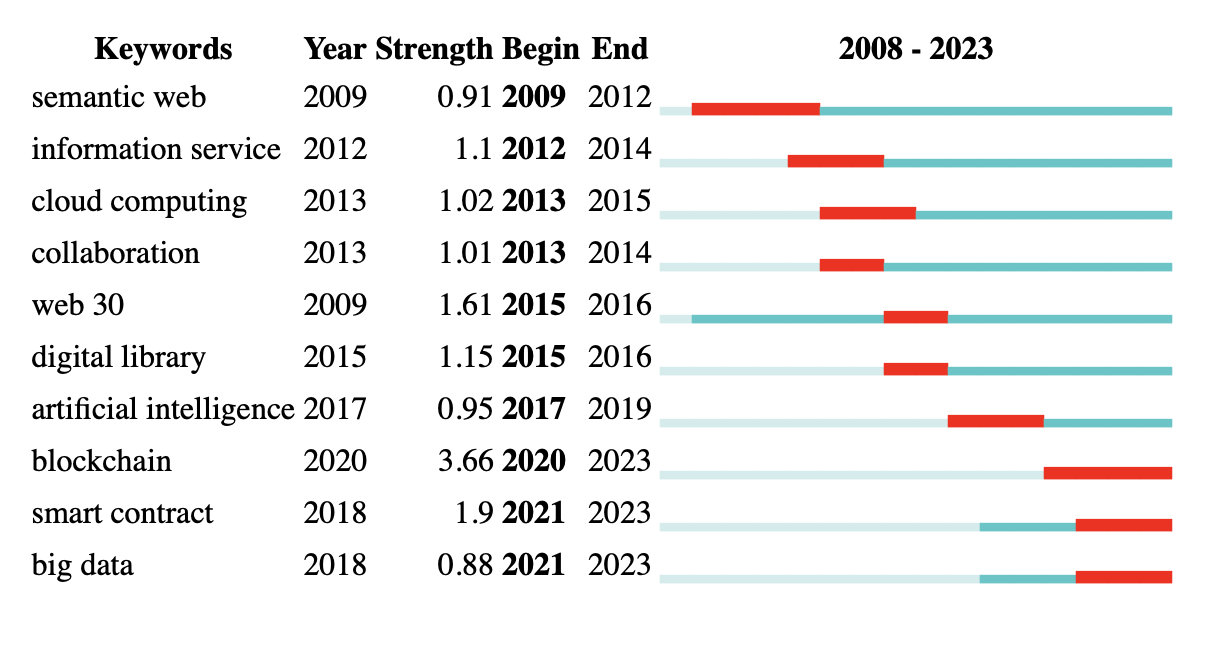} 
    \renewcommand{\figurename}{Figure}
    \caption{WOS emergent keywords Top10: light green is the basic time axis, light green is the basic time axis, dark green is the time interval from the emergence of keywords to the present, and red is the time interval of the duration of keyword emergence. The keywords in the figure are sorted by the starting time of emergence}
    \label{wburst}
\end{figure}

Figure \ref{wburst} shows the top 10 prominent keywords in foreign Web 3.0-related literature. The first emergent keyword is "Semantic Web", which emerged from 2009 to 2012. This represents the initial stage of research. Foreign literature focuses on discussing the concept and meaning of Web 3.0. Like Chinese research, the keyword with the most considerable emergent strength in foreign countries is "blockchain", whose strength is 3.66. Unlike Chinese research, in recent years, foreign research focuses on the field of Web 3.0 are still concentrated on the technical topics of "blockchain", "smart contract," and "big data". At the same time, "artificial intelligence" is the common research trend in Web 3.0 in China and abroad.

%------------------------------------------%

\section{Transformation and Opportunities}

This section examines the mutualistic interplay between the construction of the Web 3.0 ecosystem and the advancement and transformation of the digital economy. Specifically, we explore how the digital economy provides fertile ground for emerging technological development, thereby promoting scientific progress and technological innovation in Web 3.0. Furthermore, this section provides a comprehensive analysis of the changing landscape and opportunities brought about by the ongoing development of Web 3.0 technology from both technological and economic perspectives.

\subsection{Transformation and Opportunities in Economy}

The concept of the digital economy can be divided into four parts \cite{chen2022digital}: digital industrialization, industry digitization, digital management, and data value. The connections between Web 3.0 in these four aspects are shown in Table \ref{经济应用}. These connections have brought about significant changes in the digital economy, creating many opportunities in this transformation.

\begin{table}
\centering
\renewcommand{\tablename}{Table}
\caption{Connections between Web 3.0 and digital economy}
\Large
\scalebox{0.6}{
\begin{tabular}{|c|c|c|c|}
\hline
 & \textbf{Blockchain} & \textbf{Decentralization} & \textbf{Metaverse} \\ \hline
\textbf{Digital industrialization} & Digital currency and NFT & Defi and storage technology & Metaverse software \\ \hline
\textbf{Industry digitization} & IoT & E2E Mechanism & VR/AR/MR \\ \hline
\textbf{Digital management} & Smart cities & Decentralized management & - \\ \hline
\textbf{Data value} & Federated learning & Data storage & Metaverse and AI\\ \hline
\end{tabular}}
\label{经济应用}
\end{table}

From the perspective of digital industrialization, the transformation of the economic landscape by Web 3.0 is mainly reflected in value creation centered around users by delegating digital assets that initially belonged to the platform to users. Digital currency, NFT, decentralized finance and storage, and metaverse software are all specific applications. These applications have effectively promoted the development of the digital economy. For example, Web 3.0 protects digital art works and artists' rights and interests through payment, intellectual property management, digital storage, etc., which significantly promotes the creation and transaction of digital works of art in Web 3.0 \cite{cunningham2019research}.

Web 3.0 has promoted intelligence development and effective management in traditional industries regarding industry digitization. The industrial Internet of Things (IIoT) formed by Web 3.0 has been used in various industries. For example, the blockchain based on the industrial internet of things has been used in the food supply chain to ensure food quality and safety by providing a more traceable food production chain \cite{pena2019blockchain}. End-to-end technology can be described as the exchange of information between systems performing autonomous tasks, which is the foundation for the operation of future IoT devices. For the metaverse, combining Web 3.0 and VR or AR technology can make our interactions more immersive, effectively promoting the development of teaching software and electronic games.

In terms of digital management, intelligent cities utilize technology and data to improve the quality of life of their residents, enhance sustainability, and simplify urban services. The concept involves integrating information and communication technology and IoT devices into urban infrastructure to collect and analyze real-time data. Then use this data to manage urban assets and resources, provide better services to its residents, and make wise decisions about the city's future development. Decentralized management is an essential component of innovative city governance, enabling people to coordinate and govern themselves through automated execution rules deployed on the blockchain \cite{hassan2021decentralized}.

In mining data value, it is necessary to ensure both the security of the data and sufficient storage space. Among them, ensuring secure privacy computing can prevent data leakage and ensure that sensitive information is not accessed or manipulated by unauthorized persons while ensuring that organizations or individuals can process and use this data. It includes many privacy protection technologies, such as secure multi-party computing, homomorphic encryption, differential privacy, and zero-knowledge proof \cite{yang2019federated}. In addition, decentralized data storage effectively increases data transmission flexibility and scattered storage resources. AI can also drive these data to enhance participants' immersive interaction experience in the metaverse.

\subsection{Transformation and Opportunities in technology}

\begin{figure}[H]
    \centering
    \includegraphics[scale=0.5]{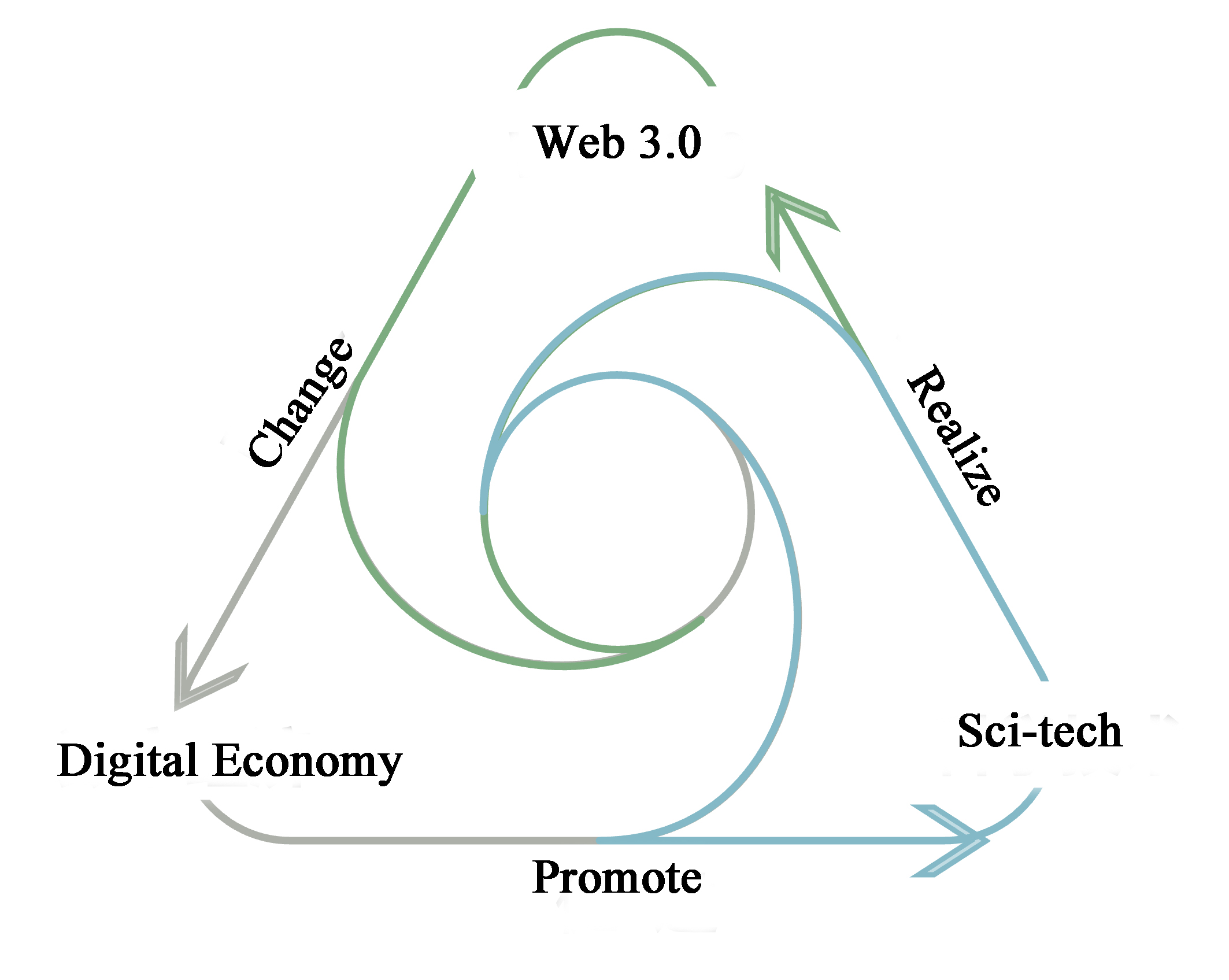} 
    \renewcommand{\figurename}{Figure}
    \caption{The Relationship between Web 3.0, Digital Economy, and Science and Technology}
    \label{WEBr}
\end{figure}

Web 3.0 requires the support of many emerging technologies, mainly AI, blockchain, and VR/AR. They all have very high requirements for computing power, requiring fundamental network technologies such as computing, transmission, storage, and related equipment to undergo leapfrog improvements. The transformation and benefits of Web 3.0 on the economy will further promote the development of related technologies. The relationships between Web 3.0, digital economy, and science and technology are shown in \ref{WEBr}

In recent years, investors have invested in the development of science and technology related to Web 3.0, and major companies have also formed new technology projects related to Web 3.0, ushering in new changes and opportunities for technological development. From the perspective of technology projects that have already been implemented, Google Cloud has launched a fully hosted cloud service that maintains blockchain nodes, known as the Blockchain Node Engine. It is mainly responsible for managing and maintaining the integrity of the blockchain ledger and promoting the consensus process among network participants. This engine can significantly reduce the cost of data management and maintenance and improve the efficiency of Web 3.0 development. In addition, WIMI.US has developed a human-machine interaction system based on XR technology, and its VR human-machine interaction terminal and other human-machine interaction-related technologies have also obtained corresponding patents.

\begin{figure}[H]
    \centering
    \includegraphics[scale=0.5]{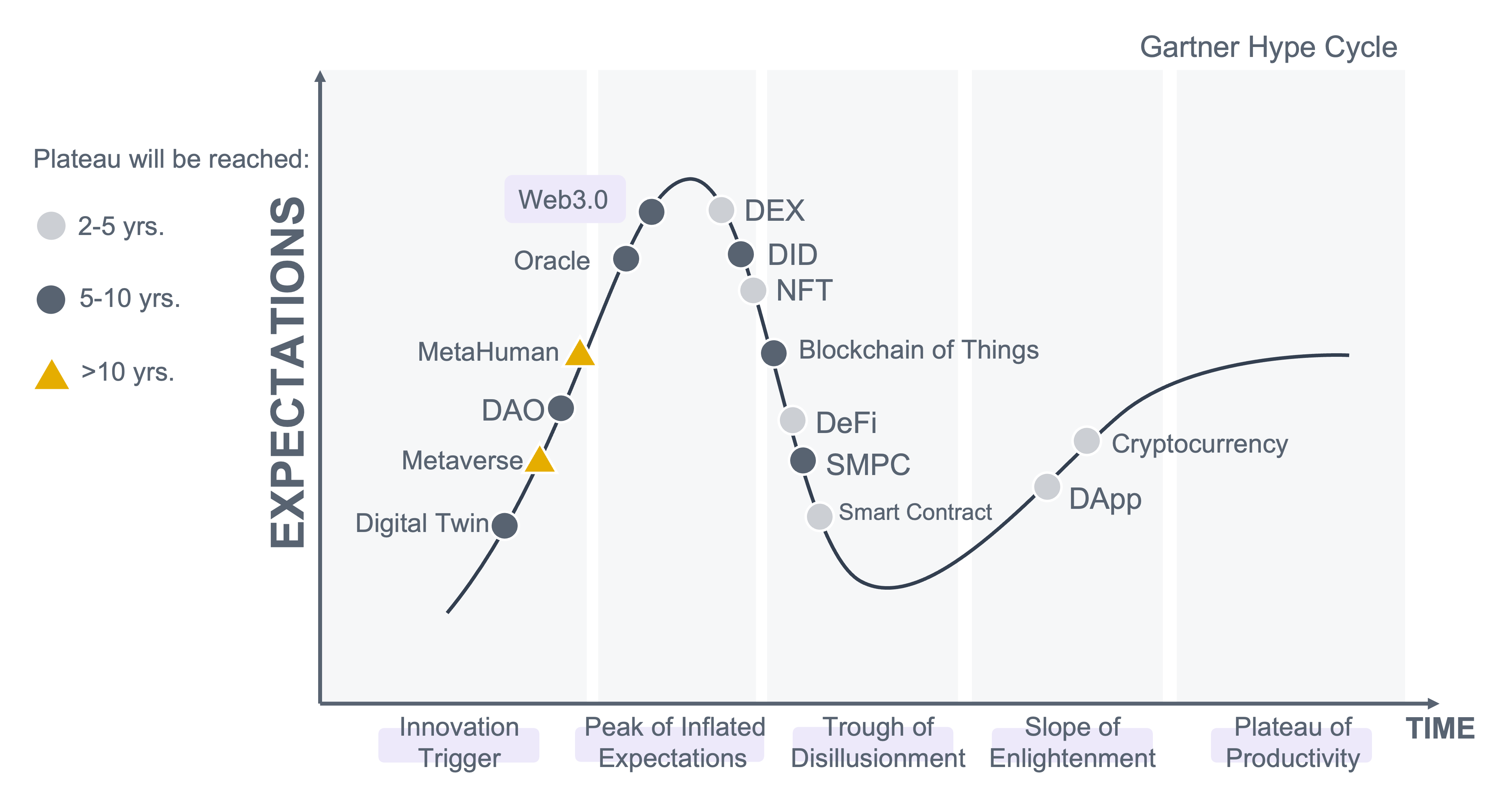}
    \renewcommand{\figurename}{Figure}
    \caption{Gartner}
    \label{gartner}
\end{figure}

Figure \ref{gartner} shows the Gartner hype cycle which include five stages of new technology would go through from inception to maturity. In the early stages of new technology, the market will have high expectations for the benefits it can bring. However, the market will eliminate unrealistic and hyped technology after the technology has gone through the foam period. Afterward, emerging technologies will gradually transition to a mature stage. At this stage, the market's expectations for new technologies will become more stable and closer to reality.

The concept of Web 3.0 first appeared in 2006, but before that, there had been a long period of decentralized and distributed system technology. As a result, most Web 3.0 applications are experimental and have yet to be widely adopted.

Blockchain technologies, such as Ethereum, have developed into mature platforms in the past few years, and many Web 3.0 applications have emerged. These applications include decentralized exchanges, digital identity verification systems, digital asset management platforms, etc. At this stage, Web 3.0 applications are being promoted to many users, and some larger companies and organizations are exploring the potential of these technologies.

However, Web 3.0-related technologies are still in the stage of innovation foam, and the current expectations for the market value are too high. False and exaggerated propaganda and hype concepts also appeared at the same time. The current market situation is relatively complex. Therefore, we need to view Web 3.0 technology more rationally, recognizing its limitations and shortcomings while facing enormous development opportunities.

%\begin{figure}[H]
%    \centering
%    \includegraphics[scale=0.9]{figures/gartner.png}
%    \renewcommand{\figurename}{Figure}
%    \caption{Gartner}
%    \label{gartner}
%\end{figure}

%------------------------------------------%

\section{Needs and Challenges}

The Web 3.0 ecosystem can solve data security, platform monopoly, and trust problems in the Web 2.0 era. At the same time, Web 3.0 provides a technical base for the metaverse, representing the direction of application scenarios and lifestyle innovation and enabling the economy to carry out digital transformation. The development of Web 3.0 is expected to shape a new system of Internet technology, optimize the development mode and industrial pattern of the Internet, and build a new paradigm of the Internet economy. The subsequent section of this manuscript will focus on identifying the key components for building a sustainable Web 3.0 ecosystem, outlining potential challenges and proposing strategies for overcoming them.

\subsection{Ecological construction}

The ecosystem of Web 3.0 is defined as a distributed open technology system that supports information and data sharing, connectivity, discovery, governance, and value aggregation. We assert that the new organizational collaboration mode, combined with advancements in financial technology and digital identity management, represents the foundational building blocks of the Web 3.0 ecology.

\subsubsection{The Management Model of Web 3.0 Ecology - Decentralized Autonomous Organizations (DAO)}

The internet ecosystem in the era of Web 2.0 is closed, as internet giants monopolize user data resources. Therefore, the public's demand for establishing a decentralized organizational collaborative approach is increasing, and DAO has also emerged. Decentralized Autonomous Organization (DAO) is a new type of organizational collaboration based on blockchain technology, which is seen as a spark that subverts traditional Internet organizational structures.

Unlike traditional organizations, DAO operates on the blockchain and is controlled by stakeholders rather than specific institutions. This allows all members to achieve co-governance within the DAO through smart contracts, eliminating subordinate relationships and improving the efficiency of organizational collaboration. Although the concept of DAO is developing in various forms, security and legal risks, remain a concern. Additionally, further research is needed to establish a DAO ecosystem that aligns with China's national conditions.

\subsubsection{The driving force of the Web 3.0 ecosystem - Decentralized Finance (DeFi)}

Decentralized Finance (DeFi) is a new type of digital finance based on blockchain technology. DeFi aims to achieve secure financial services without the intervention of traditional financial institutions through smart contracts and decentralized networks.

This new financial ecosystem is built on public chains such as Ethereum, supporting various financial applications and services, including decentralized exchanges (DEX), lending platforms, stable currencies, and unmanaged wallets. In addition, smart contracts on the blockchain can automatically execute DeFi transactions, and each transaction data is auditable. The main financial applications implemented by DeFi include asset trading, available lending, and aggregated income wealth management.

Table \ref{defi} compares the differences between decentralized finance, centralized finance, and traditional financial systems in different financial services categories. The asset trading of the crypto financial system on the chain is done through crypto exchanges, but the difference is that DeFi's exchange is decentralized. Decentralized exchanges (DEXs), as alternative payment ecosystems with new financial trading protocols, appear within the framework of available finance and are part of blockchain technology and financial technology. Unlike centralized cryptocurrency exchanges (CEX) such as Coinbase, Huobi, or Binance, they used to order books to match buyers and sellers in the open market and store encrypted assets in exchange-based wallets. In addition, DEX is unmanaged and utilizes automated intelligent contracts for peer-to-peer trading, while users retain control over their private keys and funds\cite{ref10}.

\begin{table}
\centering\vspace{-0.2cm}
\renewcommand{\tablename}{Table}
\caption{DeFi,CeFi and traditional finance}
\setlength{\tabcolsep}{1mm}{
\large
\scalebox{0.7}{
\begin{tabular}{|c|c|cc|c|}
\hline
 & & \multicolumn{2}{c|}{Crypto financial system} &
   \\ \cline{3-4}

\multirow{-2}{*}{Category} &
  \multirow{-2}{*}{Service} &
  \multicolumn{1}{c|}{\begin{tabular}[c]{@{}c@{}}Decentralized Finance\\ (DeFi) \end{tabular}} &
  {\begin{tabular}[c]{@{}c@{}}Centralized Finance\\ (CeFi) \end{tabular}} &
  \multirow{-2}{*}{Traditional Finance} \\ \hline
 & Fund transfer &
  \multicolumn{1}{c|}{\begin{tabular}[c]{@{}c@{}}DeFi Stable coin\\ (DAI) \end{tabular}} &
  \begin{tabular}[c]{@{}c@{}}CeFi Stable coin\\ (USDT,USDC) \end{tabular} &
  Payment platforms \\ \cline{2-5} 
 &
  Asset transactions  &
  \multicolumn{1}{c|}{\begin{tabular}[c]{@{}c@{}}Crypto assets DEX\\ Uniswap\end{tabular}} &
   &
   \\ \cline{2-3}
\multirow{-5}{*}{Trading} &
  Derivative trading &
  \multicolumn{1}{c|}{\begin{tabular}[c]{@{}c@{}}Crypto derivative DEX\\ Synthetic,dYdX\end{tabular}} &
  \multirow{-3}{*}{\begin{tabular}[c]{@{}c@{}}Crypto CEX\\(Binance,Coinbase)\end{tabular}} &
  \multirow{-3}{*}{\begin{tabular}[c]{@{}c@{}}Bourse\\Agent \end{tabular}} \\ \hline
 &
  Mortgage &
  \multicolumn{1}{c|}{\begin{tabular}[c]{@{}c@{}}Decentralized lending platform\\(Aave,Compound)\end{tabular}} &
  \begin{tabular}[c]{@{}c@{}}Centralized lending platform\\ (BlockFi,Celsius)\end{tabular} &
  Loan Trading agents \\ \cline{2-5} 
\multirow{-3}{*}{Debit} &
  Unsecured Loan &
  \multicolumn{1}{c|}{\begin{tabular}[c]{@{}c@{}}Credit commission agency\\(Aave)\end{tabular}} &
  \begin{tabular}[c]{@{}c@{}}Digital banking\\ (Silvergate)\end{tabular} &
  Commercial bank  \\ \hline
Investment &
  Financial Products  &
  \multicolumn{1}{c|}{\begin{tabular}[c]{@{}c@{}}Decentralized investment portfolio\\(yearn,Convex)\end{tabular}} &
  \begin{tabular}[c]{@{}c@{}}Crypto Fund
\\(Grayscale,Galaxy)\end{tabular} &
  Fund \\ \hline
\end{tabular}}}
\label{defi}
\end{table}

From the table above, DeFi is also the digitized form of the traditional financial system. It cannot operate independently from the traditional financial system or avoid the hidden risks in traditional finance. However, DeFi has the advantage of having high transparency and coverage that traditional finance cannot possess.

\subsubsection{Distributed digital identity (DID): an ecological identity authentication method for Web 3.0}

The emergence of distributed digital identity (DID) has solved the problem of self-sovereignty of digital identity in networking and enhanced the individual's control over identification information. Specifically, DID aims to form a self-sovereignty identity where users can exercise autonomous control over their data information without needing permission from any platform or organization. DID is a critical component of the constantly evolving Web 3.0 ecosystem, which can verify identity without relying on centralized institutions and third-party services, providing a method that does not require trust and guarantees privacy.

\subsubsection{Industrial ecological value}

The Web 3.0 ecosystem is linked by token incentives, which can significantly achieve co-construction and sharing within organizations, break boundaries, and promote value circulation. First, underpinning the Web 3.0 ecosystem is a protocol that somewhat reduces the cost of obtaining trust guarantees from third-party sources, offering individuals more opportunities to exchange their innovations for monetary gains directly. Within the decentralized autonomous organization (DAO), participants can obtain benefits transparently and with greater trust through the interconnected distributed digital identity.

Secondly, the Web 3.0 ecosystem can transcend industry boundaries to create new forms of collaboration. The enhanced digital system integrated with advanced technologies in Web 3.0 can efficiently document, monitor, and adjust every aspect of the interaction between connected devices. In this more open internet environment, blockchain-based digital trust systems expand the coverage of digital services and accelerate data integration into reality, ultimately nurturing low-cost, efficient, and healthier socio-economic ecosystems.

As such, token incentives are crucial for establishing and maintaining value circulation within the Web 3.0 ecosystem, creating a highly interoperable and trusted environment that enhances industrial cooperation across sectors. Nonetheless, there is still much work required to align the adoption and development of Web 3.0 with China's national policies while positively contributing to global digital innovation initiatives.

\subsection{Current challenges}

Although the construction of the Web 3.0 ecosystem has become a vision for the development of the Internet, it still faces many challenges.

Primarily, there is a lack of a systematic theoretical and technological research layout for Web 3.0 in China. Research into Web 3.0 theory and technology, such as architecture and protocol stacks, alliance chains, open alliance chains, public chains, digital identities, and distributed storage, requires significant backing from artificial intelligence, cloud computing, 5G, virtual reality, and other emerging technologies.However, while there has been more research focus on the alliance chain in China, robust research initiatives are still lacking on these other critical components. Another critical challenge associated with Web 3.0's development is the severe shortage of high-level, composite talents in this field. The complexity and diversity involved in Web 3.0's technology and application demand proficiency and cross-disciplinary knowledge in cryptography, computer science, network security, visualization and economics\cite{Cheng2016,Zhang2022}.Despite this, there is still no specialized course or major to cultivate Web 3.0 talents in China specifically. That, coupled with the lack of related textbooks, training platforms, and other supportive resources, indicates an urgent need to establish a more comprehensive talent cultivation system that can stay abreast of the rapidly evolving field of Web 3.0.

Secondly,The decentralized characteristics of Web 3.0 may pose significant challenges to existing legal and regulatory systems. For instance, some questions need to be addressed on how to enhance ownership relationships associated with NFTs and develop a property management system to govern the legal aspects of NFT ownership.Additionally, the emergence of new business models facilitated by Web 3.0 may introduce illegal and criminal activities, such as illegal fundraising and money laundering, which could disrupt the financial market. Legal frameworks must be put in place to regulate these activities effectively.Another significant challenge pertains to identifying DAOs' nature from a legal perspective, assigning appropriate rights and obligations, and establishing regulations that guide their operations. It is noteworthy that public chains and cryptocurrencies play an integral role in the development and success of Web 3.0. However, due to limited development by domestic regulatory policies, scholars are keen to explore ways to circumvent these limitations.As such, it is essential to address these legal challenges comprehensively to ensure the successful development and deployment of Web 3.0 while safeguarding the public interest.

\section{Conclusion}

Web 3.0 is still in its early stages of development and has enormous potential for development. With the development of science and technology such as 5G, AI, and metaverse, as well as the further improvement of management concepts such as decentralization, Web 3.0 will make breakthroughs in the following two areas in the future, allowing users to receive better services \cite{rudman2016defining}. 1) With the further improvement of decentralized data management and the 5G, the workload of data on a single server will be effectively reduced, and the existing data transmission efficiency will be further improved to more efficiently search or analyze data for users. 2) With the further optimization of AI technology, they will be better integrated with Web 3.0 and provide intuitive web services to promote economic housing exhibitions. For example, AI can summarize many consumer habits collected, leading to better online marketing and promoting consumption. Combining efficient data management and AI will bring users more valuable and personalized data services. In the future, more AI platforms similar to ChatGPT will provide users with accurate and efficient data search and generation services \cite{ouyang2022training,huang2023roadmap}.

It is for sure that potential risks would appear along with the development of Web 3.0, specifically regarding legal and technical factors. Concerning decentralized data management, challenges may emerge in regulating the proliferation of illicit data on the network, as well as enhancing cybersecurity to managing network attacks like SQL injection and malware, which are more likely in the Web 3.0 environment. Personalizing network content may also raise concerns regarding personal data privacy and the associated risks of identity theft or social phishing.

Given these prospects and challenges, China must examine its governance policies and regulatory frameworks surrounding Web 3.0 while simultaneously focusing on cultivating requisite technology capabilities and supporting infrastructural improvements to ensure effective management of the decentralized networks.

Through this approach, China can shape a distinctive model for Web 3.0 development with reinforced central governance mechanisms and robust legislation that better safeguards user safety and privacy and maintains national security. As such, it is arguable that China possesses relative strategic advantages given its unique geopolitical position, significant technological capacity, and ambitions toward technological leadership. Overall, exploring avenues for effective governance frameworks that balance innovation and business opportunities with sustainable risk management will be critical to successfully realizing Web 3.0's promise and potential.

%------------------------------------------%
\newpage 
\bibliography{bibliography}
\bibliographystyle{plain}
%\printbibliography[heading=bibintoc,title={References}]

%\end{CJK}
\end{document}